\input harvmac
\input epsf
\input slashed.sty

\def\M{{\cal M}}
\def\W{{\cal W}}
\def\O{{\cal O}}
\def\T{{\cal T}}
\def\D#1{{{d^d#1}\over{(2\pi)^d}}}
\def\frac#1#2{{{#1}\over{#2}}}
\def\Li{\hbox{Li}_2}
\def\vev#1{\langle#1\rangle}
\def\\#1 {{\tt\char'134#1} }
\def\centertable#1{ \hbox to \hsize {\hfill\vbox{\offinterlineskip \tabskip=0pt \halign{#1} }\hfill} }
\def\wt{\widetilde}
\lref\GiudiceBP{
  G.~F.~Giudice and R.~Rattazzi,
  ``Theories with gauge-mediated supersymmetry breaking,''
  Phys.\ Rept.\  {\bf 322}, 419 (1999)
  [arXiv:hep-ph/9801271].
}
\lref\GiudiceNI{
  G.~F.~Giudice and R.~Rattazzi,
  ``Extracting supersymmetry-breaking effects from wave-function renormalization,''
  Nucl.\ Phys.\  B {\bf 511}, 25 (1998)
  [arXiv:hep-ph/9706540].
}
\lref\GM{
  S.~Dimopoulos and S.~Raby,
  ``Supercolor,''
  Nucl.\ Phys.\  B {\bf 192}, 353 (1981)\semi
  M.~Dine and W.~Fischler,
  ``A Phenomenological Model Of Particle Physics Based On Supersymmetry,''
  Phys.\ Lett.\  B {\bf 110}, 227 (1982)\semi
  C.~R.~Nappi and B.~A.~Ovrut,
  ``Supersymmetric Extension Of The SU(3) X SU(2) X U(1) Model,''
  Phys.\ Lett.\  B {\bf 113}, 175 (1982)\semi
  L.~Alvarez-Gaume, M.~Claudson and M.~B.~Wise,
  ``Low-Energy Supersymmetry,''
  Nucl.\ Phys.\  B {\bf 207}, 96 (1982)\semi
  M.~Dine and A.~E.~Nelson,
  ``Dynamical supersymmetry breaking at low-energies,''
  Phys.\ Rev.\  D {\bf 48}, 1277 (1993)
  [arXiv:hep-ph/9303230]\semi
  M.~Dine, A.~E.~Nelson and Y.~Shirman,
  ``Low-Energy Dynamical Supersymmetry Breaking Simplified,''
  Phys.\ Rev.\  D {\bf 51}, 1362 (1995)
  [arXiv:hep-ph/9408384]\semi
  M.~Dine, A.~E.~Nelson, Y.~Nir and Y.~Shirman,
  ``New tools for low-energy dynamical supersymmetry breaking,''
  Phys.\ Rev.\  D {\bf 53}, 2658 (1996)
  [arXiv:hep-ph/9507378].
}
\lref\UoneA{
  H.~C.~Cheng, B.~A.~Dobrescu and K.~T.~Matchev,
  ``A chiral supersymmetric standard model,''
  Phys.\ Lett.\  B {\bf 439}, 301 (1998)
  [arXiv:hep-ph/9807246].
}
\lref\UoneB{
  H.~C.~Cheng, B.~A.~Dobrescu and K.~T.~Matchev,
  ``Generic and chiral extensions of the supersymmetric standard model,''
  Nucl.\ Phys.\  B {\bf 543}, 47 (1999)
  [arXiv:hep-ph/9811316].
}
\lref\UoneC{
  D.~E.~Kaplan, F.~Lepeintre, A.~Masiero, A.~E.~Nelson and A.~Riotto,
  ``Fermion masses and gauge mediated supersymmetry breaking from a single U(1),''
  Phys.\ Rev.\  D {\bf 60}, 055003 (1999)
  [arXiv:hep-ph/9806430].
}
\lref\UoneD{ 
  L.~L.~Everett, P.~Langacker, M.~Plumacher and J.~Wang,
  ``Alternative supersymmetric spectra,''
  Phys.\ Lett.\  B {\bf 477}, 233 (2000)
  [arXiv:hep-ph/0001073].
}
\lref\Zprime{
  P.~G.~Langacker, G.~Paz, L.~T.~Wang and I.~Yavin,
  ``Z'-mediated Supersymmetry Breaking,''
  arXiv:0710.1632 [hep-ph].
}
\lref\MartinZB{
  S.~P.~Martin,
  ``Generalized messengers of supersymmetry breaking and the sparticle mass spectrum,''
  Phys.\ Rev.\  D {\bf 55}, 3177 (1997)
  [arXiv:hep-ph/9608224].
}
\lref\DimopoulosGY{
  S.~Dimopoulos, G.~F.~Giudice and A.~Pomarol,
  ``Dark matter in theories of gauge-mediated supersymmetry breaking,''
  Phys.\ Lett.\  B {\bf 389}, 37 (1996)
  [arXiv:hep-ph/9607225].
}
\lref\IntriligatorDD{
  K.~Intriligator, N.~Seiberg and D.~Shih,
  ``Dynamical SUSY breaking in meta-stable vacua,''
  JHEP {\bf 0604}, 021 (2006)
  [arXiv:hep-th/0602239].
}
\lref\vanderBijBW{
  J.~van der Bij and M.~J.~G.~Veltman,
  ``Two Loop Large Higgs Mass Correction To The Rho Parameter,''
  Nucl.\ Phys.\  B {\bf 231}, 205 (1984).
}
\lref\MartinZK{
  S.~P.~Martin and M.~T.~Vaughn,
  ``Two Loop Renormalization Group Equations For Soft Supersymmetry Breaking Couplings,''
  Phys.\ Rev.\  D {\bf 50}, 2282 (1994)
  [arXiv:hep-ph/9311340].
}
\lref\ArkaniHamedKJ{
  N.~Arkani-Hamed, G.~F.~Giudice, M.~A.~Luty and R.~Rattazzi,
  ``Supersymmetry-breaking loops from analytic continuation into  superspace,''
  Phys.\ Rev.\  D {\bf 58}, 115005 (1998)
  [arXiv:hep-ph/9803290].
}
\lref\MartinVX{
  S.~P.~Martin,
  ``Two-loop effective potential for a general renormalizable theory and
  softly broken supersymmetry,''
  Phys.\ Rev.\  D {\bf 65}, 116003 (2002)
  [arXiv:hep-ph/0111209].
}
\lref\PoppitzXW{
  E.~Poppitz and S.~P.~Trivedi,
  ``Some remarks on gauge-mediated supersymmetry breaking,''
  Phys.\ Lett.\  B {\bf 401}, 38 (1997)
  [arXiv:hep-ph/9703246].
}
\lref\SiegelWQ{
  W.~Siegel,
  ``Supersymmetric Dimensional Regularization Via Dimensional Reduction,''
  Phys.\ Lett.\  B {\bf 84}, 193 (1979).
}
\lref\DermisekQJ{
  R.~Dermisek, H.~D.~Kim and I.~W.~Kim,
  ``Mediation of supersymmetry breaking in gauge messenger models,''
  JHEP {\bf 0610}, 001 (2006)
  [arXiv:hep-ph/0607169].
}
\lref\KaplunovskyYX{
  V.~Kaplunovsky,
  ``Nosonomy Of An Upside Down Hierarchy Model. 2,''
  Nucl.\ Phys.\  B {\bf 233}, 336 (1984).
}
\lref\ChengYS{
  H.~C.~Cheng and D.~E.~Kaplan,
  ``Axions and a gauged Peccei-Quinn symmetry,''
  arXiv:hep-ph/0103346.
}
\lref\CheungES{
  C.~Cheung, A.~L.~Fitzpatrick and D.~Shih,
  ``(Extra)Ordinary Gauge Mediation,''
  arXiv:0710.3585 [hep-ph].
}
\lref\LangackerYV{
  P.~Langacker,
  ``The Physics of Heavy Z' Gauge Bosons,''
  arXiv:0801.1345 [hep-ph].
}
\lref\BaggerBT{
  J.~A.~Bagger, K.~T.~Matchev, D.~M.~Pierce and R.~j.~Zhang,
  Phys.\ Rev.\  D {\bf 55}, 3188 (1997)
  [arXiv:hep-ph/9609444].
}
\lref\HillerQG{
  G.~Hiller and M.~Schmaltz,
  Phys.\ Lett.\  B {\bf 514}, 263 (2001)
  [arXiv:hep-ph/0105254].
}
\Title{UCSD-PTH-08-02}{Sparticle Masses in Higgsed Gauge Mediation}

\centerline{Elie Gorbatov and Matthew Sudano}
\bigskip\centerline{Department of Physics}
\centerline{University of California, San Diego}
\centerline{La Jolla, CA 92093-0319}
\vskip .3in
We generalize the gauge sector of gauge-mediated supersymmetry breaking to allow for an arbitrary gauge group with an arbitrary supersymmetric Higgsing.  The sparticle masses are computed to leading order in the gauge coupling.  The generic effect on the MSSM spectrum from additional Higgsed gauge structure is to increase the sfermion masses relative to the gaugino masses.  
\Date{February 2008} 
\baselineskip=19pt plus 2pt minus 2pt
\newsec{Introduction}
There are countless ways in which the standard model could fit into a supersymmetric framework.  For the sake of phenomenology, a useful categorization is in terms of how supersymmetry breaking is communicated to the observable particles.  Gauge-mediated supersymmetry breaking (GMSB) \GM\ assumes that SUSY breaking is communicated via the standard model gauge group.  This mechanism has several attractive features.  In particular, it makes calculable predictions for the soft parameters of the MSSM in terms of a few parameters while naturally evading the tight constraints from flavor physics.  It also accommodates radiative electroweak symmetry breaking \BaggerBT\ and offers a solution to the CP problem \HillerQG.  

It is important to remember, however, that this is not a complete theory.  It is only meant to apply below the scale of SUSY-breaking.  The hope is that it provides a successful parameterization of our ignorance of physics at the higher scale, but this has recently been called into question.  In \CheungES, for example, it was pointed out that the standard approach omits a set of renormalizable interactions that are allowed by the symmetries, are consistent with experiment, and lead to novel phenomenology.  In this note, we explore another generalization of ordinary gauge mediation that has previously been ignored.  Specifically, we consider a general, supersymmetric Higgsing of the mediating gauge group.

In a sense, this is not new at all because it is generally assumed that there is a supersymmetric 
Higgsing of the mediating gauge group, both at the GUT scale and at the weak scale.  For a messenger scale much larger than the weak scale, the masses of the $SU(2)_W$ gauge fields can be neglected.  And the gauge fields with GUT-scale masses can be ignored if the messenger scale is sufficiently small, but for models with a messenger scale near the GUT scale, as in \DermisekQJ, they can be very important.   Of course, gauge symmetry breaking may also occur at an intermediate scale.  For example, additional $U(1)$'s \refs{\Zprime\UoneA\UoneB\UoneC\UoneD{--}\LangackerYV}, which arise naturally in SUSY-GUTs with large gauge groups and in string theory, typically have intermediate Higgsing scales. 
\centerline{\epsfxsize=2.55in\epsfbox{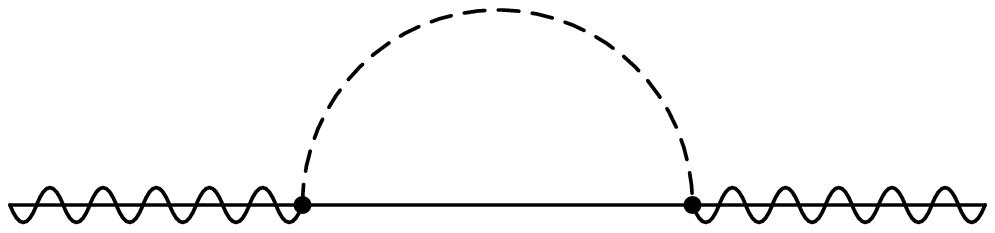}}
\centerline{Figure 1: The sole one-loop diagram contributing to gaugino masses}
\vskip .2in
In what follows, we will briefly review the standard treatment of GMSB and the associated sparticle spectrum.  We will then discuss how one can approximate these results for much of parameter space, and we will show that these techniques fail to capture the effects of interest here.  Finally, we will present the leading-order sparticle spectrum in standard GMSB for an arbitrary, supersymmetric Higgsing of the gauge group, and comment on the results.  The messier details of the calculation are left for the appendix.  
\newsec{Standard Gauge Mediation}
In the basic scenario (see \GiudiceBP \ for a review), a set of chiral superfields, $\Phi_i$ and $\tilde\Phi_i$, are added to a GUT-extended MSSM.  They can all be taken to be {\bf 5} and $\bar{\bf5}$ of $SU(5)$, for example.  Note that this choice preserves gauge-coupling unification, is anomaly-free, and allows for the superpotential term, 
\eqn\w{\Delta W=\sum_i\lambda_{i}X\tilde\Phi_i\Phi_i,}
where $X=M+F\theta^2$ is a gauge-singlet background field.  The index, $i$, is a flavor index.  Gauge indices are supressed.  The non-zero F-component of the ``spurion'', $X$, results in a non-supersymmetric mass spectrum for our chiral superfields, which then act as messengers, splitting masses of MSSM superfields through loops.   Their contributions to MSSM gaugino and scalar masses have been computed for the scenario described above and for some generalizations \refs{\DimopoulosGY,\MartinZB, \PoppitzXW}\footnote{$^1$}{In principle, the two-loop effective potential of \MartinVX\ contains these scalar masses and those presented later in this paper.  In practice, however, extracting such results is difficult with basic computational resources in anything but the simplest theories.}.  The gauginos get masses,
\eject
\centerline{\epsfxsize=2.55in\epsfbox{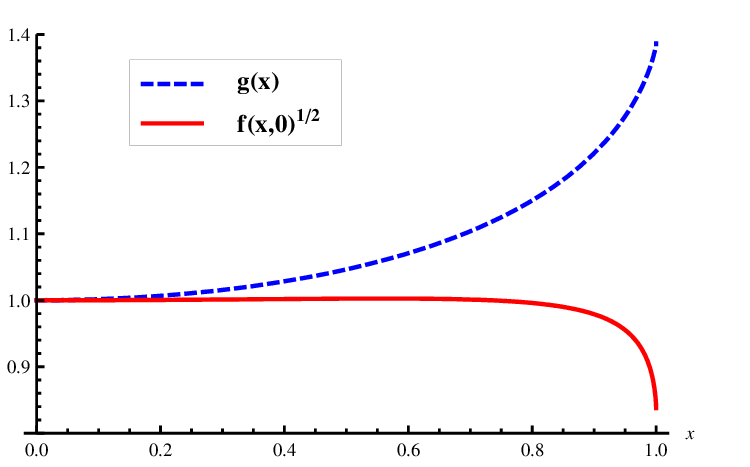}}
\centerline{Figure 2: $f(0,0)=1=g(0)$, but $g(1)/f(1,0)^{1/2}\approx5/3$}
\vskip.2in
\eqn\gm{\Delta m_{1/2}^a=\frac{\alpha_a}{2\pi}\frac{F}{M}\sum_in_a(r_i)g(x_i),\qquad x_i=\Big|\frac{F}{\lambda_iM^2}\Big|,}
\noindent from the diagram in Figure 1, and the first eight diagrams of Figure 5 give the scalar masses,
\eqn\sm{\Delta m_0^2=\Big|\frac{F}{M}\Big|^2\sum_a\Big(\frac{\alpha_a}{2\pi}\Big)^2C_a(r_Q)\sum_in_a(r_i)f(x_i,0).}
Much of the notation is the same as that of \MartinZB.  The index,  $a=1,2,3$, labels the gauge group,   $\alpha_a=g_a^2/4\pi$, $C_a(r_Q)$ is the quadratic Casimir of the scalar field that is getting mass, and  $n_a(r_i)$ is the Dynkin index of the messenger representation.  The extra argument in the function, $f(x_i,0)$, will be explained shortly.  

Note that \gm\ is a mass, and \sm\ is a mass squared, so in the ratio of a gaugino and a scalar mass, the factors of $F/M$ cancel.  And as one can see in Figure 2, the functions, $g(x)$ and $f(x,0)^{1/2}$, deviate little from one over most of parameter space.\footnote{$^2$}{Note that $x_i$ cannot exceed one.  This would give a tachyonic messenger, and there is nothing to stabilize the field.  In general, however, the UV completion can accommodate $x>1$.  The negative mass-squared simply indicates that the true vacuum is elsewhere, and in that vacuum, there is a massive gauge field.}  This means that $m_{1/2}^a/m_0$ primarily depends on the ``effective messenger number'', $N_a\equiv2\sum_i n_a(r_i)$.  With the conventional normalization of the generators, $n_a=1/2$ for fundamentals, so in \eject 
\noindent the simple case of $SU(5)$ with fundamental messengers, the effective messenger number is the number of messengers.  This is one way in which measuring only a couple of soft parameters of the MSSM could reveal something about the messenger sector.  We will see, however, that this simple picture can be modified when the mediating gauge group is Higgsed.  
\newsec{Analytic Continuation to Superspace}
In the limit of small supersymmetry breaking, the results of the previous section can be obtained in an entirely different way \refs{\KaplunovskyYX,\GiudiceNI,\ArkaniHamedKJ}.  
Consider a massless chiral superfield, $Q$, that only couples to the messengers through gauge fields.  The Lagrangian will have a term,
\eqn\L{{\cal L}\supset\int d^4\theta Z_Q(\mu)Q^\dagger Q,}
where $Z_Q(\mu)$ is the wave-function renormalization of $Q$ at the scale $\mu$.  If this scale is below the messenger scale, then $Z_Q(\mu)=Z_Q(\mu,M^\dagger,M)$.    
Now comes the interesting part.  The idea is to replace $M$ with the superfield, $X=M+F\theta^2$.  This new object, call it $\tilde Z_Q(\mu)$, has an expansion in powers of $\theta$, which yields a mass for the scalars, 
\eqn\smac{\Delta m_0^2(\mu)=-\bigg|\frac{F}{M}\bigg|^2\frac{\partial^2\ln\tilde Z_Q(\mu)}{\partial\ln X\partial\ln X^\dagger}\bigg|_{X=M}.}
Performing the derivatives, one finds agreement with \sm\ to ${\cal O}(x^2)$ for $x=F/M^2\ll1$.  

What we are interested in is the spectrum when we have chiral messengers and a supersymmetric Higgsing.  If we take $\Lambda_{UV}>M>m_W>\mu$, then we have
\eqn\Zsbgm{Z_Q(\mu)=Z_Q(\Lambda_{UV})\Big(\frac{\alpha(\Lambda_{UV})}{\alpha(M)}\Big)^{2C(r_Q)/b}\Big(\frac{\alpha(M)}{\alpha(m_W)}\Big)^{2C(r_Q)/b'}\Big(\frac{\alpha(m_W)}{\alpha(\mu)}\Big)^{2C'(r_Q)/b''} 
$$
$$
\alpha^{-1}(M)=\alpha^{-1}(\Lambda_{UV})+\frac{b}{4\pi}\ln\frac{M^\dagger M}{\Lambda_{UV}^2},
$$
$$
\alpha^{-1}(m_W)=\alpha^{-1}(M)+\frac{b'}{4\pi}\ln\frac{m_W^2}{M^\dagger M}
$$
$$
\alpha^{-1}(\mu)=\alpha^{-1}(m_W)+\frac{b''}{4\pi}\ln\frac{\mu^2}{m_W^2}.}
Making the substitution, $M\rightarrow X$, and plugging into \smac\ gives a mass that depends on $m_W$, but only trivially.  It only acts to give the correct running of the coupling to the scale, $\mu$.  This should not be surprising since the method takes the gauge fields to be massless above $m_W$ and infinitely massive below.  In a sense, what we are interested in is a threshold effect.  The gauge messenger case, in which the spurion breaks the gauge group, is different because the scale, $m_W=M$, enters in the Grassman-parameter expansion.  Perhaps there is a clever way of approximating a non-trivial effect of a supersymmetric Higgsing, but we will not pursue this further here.  Instead, we perform the Feynman-diagram calculation. 
\newsec{Higgsed Gauge Mediation}
We are interested in the effects of modifying the gauge sector of gauge mediation.  In particular, we allow for massive gauge fields coupled to both messengers and MSSM fields, but do not study the gauge messenger scenario in which a gauge superfield has split masses.  
\subsec{Case 1 -- $G\times U(1)'$, A Toy Model}
Starting with the simplest extension, consider the set of messenger fields, $\Phi_i$ and $\tilde\Phi_i$, as in the introduction.  Now let them be charged under an additional $U(1)'$ gauge symmetry that is spontaneously broken.  The desired spectrum is obtained if we add fields, $\Psi$ and $\tilde\Psi$, that are only charged under this $U(1)'$, and take the superpotential to be
\eqn\ww{\Delta W=\sum_i\lambda_iX\tilde\Phi_i\Phi_i+hT(\tilde\Psi\Psi-v^2),\qquad X=M+F\theta^2,}
where the field, $T$, is a singlet dynamical field, which, for our purposes, plays no role except to give vevs to the added superfields.  Suppressing all indices, this produces a massive vector multiplet, ($A$, $C$, $\lambda$, $\chi$), with supersymmetric mass $m_W=2gv$, where $A$ is a gauge boson,  $C$ is a real scalar field, $\lambda$ is a gaugino, and $\chi$ is another Weyl fermion.  

Turning to the radiative spectrum, the gauginos of the unbroken gauge group get the standard one-loop masses \gm\ computed in \MartinZB, which we reproduce here:
 \centertable{
\vrule height2.75ex depth0.25ex width 0pt \tabskip=.8em  \hfil#\hfil&#&\hfil#\hfil\cr
\epsfxsize=2.5in\epsfbox{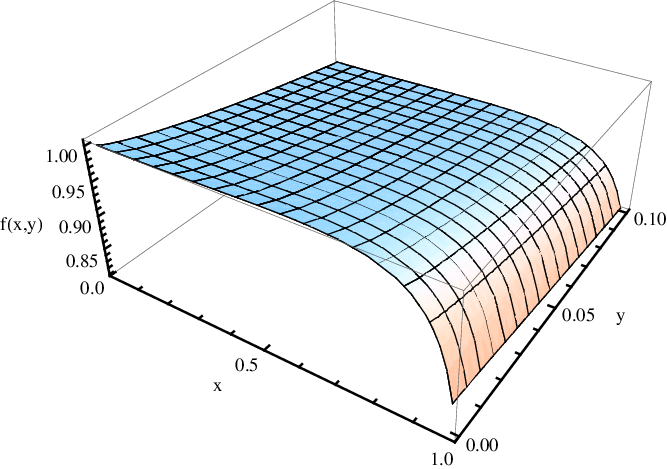}&&\epsfxsize=2.4in\epsfbox{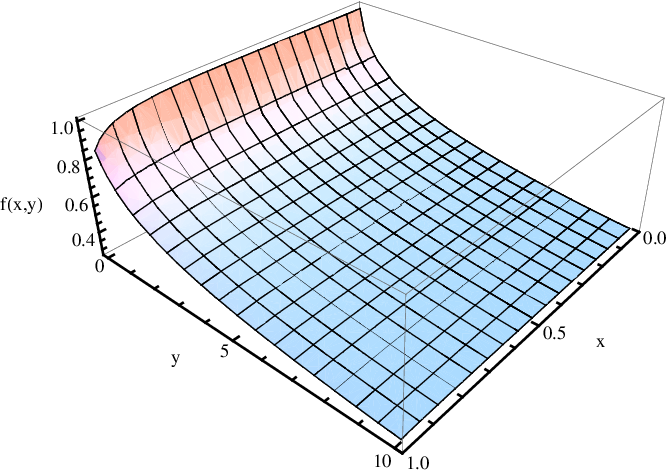}\cr
{\bf a.} && {\bf b.} \cr}
\centerline{Figure 3: $f(x,y)$ is plotted for small (a.) and large (b.) values of $y$.}
\vskip .2in
\eqn\gmass{\Delta m_{1/2}^a=\frac{\alpha_a}{2\pi}\frac{F}{M}\sum_in_a(r_i)g(x_i),}
where
\eqn\gx{g(x)=\frac{1}{x^2}(1+x)\ln(1+x)+(x\rightarrow-x).}
The notation is discussed after \sm.  To leading order, the effect of the $U(1)'$ on the gaugino spectrum is simply to add a gaugino of mass $m_W$.  The generalization to more interesting gauge structure is trivial and will not be discussed further.  
Now if we couple some set of chiral superfields, $Q$, that transform under the given gauge symmetry, their scalar components will acquire radiative masses at two-loop order.  The contribution from $G$ is as before \refs{\DimopoulosGY,\MartinZB},
\eqn\smt{\Delta m_0^2=\Big|\frac{F}{M}\Big|^2\sum_a\Big(\frac{\alpha_a}{2\pi}\Big)^2C_a(r_Q)\sum_in_a(r_i)f(x_i,0),}
where
\eqn\fxo{f(x,0)=\frac{1+x}{x^2}\bigg[\ln(1+x)-2\Li\bigg(\frac{x}{1+x}\bigg)+\half\Li\bigg(\frac{2x}{1+x}\bigg)\bigg]+(x\rightarrow-x).}
With a Higgsed mediating gauge group, there are ten relevant diagrams, which are shown in Figure 5.  For our toy model, the contribution from the $U(1)'$ vector multiplet is 
\eject
\eqn\smz{\Delta m_0^2=\Big|\frac{F}{M}\Big|^2\Big(\frac{\alpha}{2\pi}\Big)^2C(r_Q)\sum_in(r_i)f(x_i,y),\qquad y=\Big|\frac{m_W}{M}\Big|^2,} 
and $C(r_Q)$ and $n(r_i)$ are respectively the squared charges of $Q$ and $\Phi_i$ under the $U(1)'$.  The function, $f(x,y)$, is given in the appendix along with more details of the computation.  In Figure 3, this function is plotted in the limits of large and small $y$.  At small $y$, it is seen to agree with the known result of Figure 2.  At large $y$, the kinematic suppression of the amplitude is evident.  More explicitly, we find
\eqn\expansion{f(x,y\ll1)=f(x,0)+\Big(\frac{1}{3}+\frac{x^2}{30}+{\cal O}(x^4)\Big)y\ln y+{\cal O}(y)
$$
$$
f(x,y\gg1)=\frac{2}{y}\ln y+{\cal O}\Big(\frac{1}{y}\Big)}
\vskip-.2in
\subsec{Case 2 -- Products of Simple Groups with Degeneracy}
When each factor of the mediating gauge group is simple and has a single supersymmetric mass for all of its gauge superfields, the result is a simple extension of what was done in the previous section.  In fact, it is simply \smt\ with the substitution,  
\eqn\sub{f(x_i,0)\rightarrow f(x_i,y_a).}  
\vskip-.2in  
\subsec{Case 3 -- Products of Simple Groups without Degeneracy}
In generalizing to an arbitrary supersymmetric Higgsing, the most obvious obstacle is that the gauge field associated with a given generator need not be a mass eigenstate.  This is familiar from the standard model, in which the $U(1)_Y$ generator mixes with the diagonal generator of $SU(2)_W$ to form the massive $Z$ and the massless photon.  It is typically most convenient to calculate in the mass eigenbasis, working with ``effective generators'' that are linear combinations of the original generators.  With this strategy, one can quickly work out the result for a general Higgsing of a product of simple Lie groups, 
\eqn\smnd{\Delta m_0^2=\Big|\frac{F}{M}\Big|^2\sum_a\Big(\frac{\alpha_a}{2\pi}\Big)^2\sum_j\T_{a,Q}^j\T_{a,Q}^j\sum_in_a(r_i)f(x_i,y_{a,j}),\qquad y_{a,j}=\Big|\frac{\M_a^{jj}}{M}\Big|^2.}
The effective generators of each group are given by $\T^j=\O^{jk}T^k$, where $\O$ is the orthogonal matrix that diagonalizes the mass matrix of the gauge fields.  In matrix notation,
\eqn\mwd{\half W^TMW=\half W^T\O^T\O M\O^T\O W=\half(\O W)^T\M(\O W)\equiv\half\W^T\M\W,}
where $\M$ is diagonal, and $\W$ is the vector of mass eigenstates.  The effective generators emerge when the covariant derivative is written in this basis. Note that in the case of full degeneracy, $\M^{jk}=m_W\delta^{jk}$, summing over $j$ in \smnd\ reproduces the familiar quadratic Casimir, $\O^{jk}T^k\O^{jl}T^l=T^kT^l\delta^{kl}=C(r_q)$.
\subsec{Case 4 -- Allowing for $U(1)$'s}
If the gauge group includes a $U(1)$, the potential for a new complication emerges.  Fortunately, the problem and its solution are found in the simple case of a product of two $U(1)$'s.  The result for an arbitrary Higgsing of an arbitrary gauge group is easily obtained from this case; though we will not attempt to write a formula for the general case.  

Letting the gauge superfields have masses $m_W$ and $\wt m_W$ and couplings $g$ and $\tilde g$, one expects in general to have a contribution proportional to $g^2\tilde g^2$.  The presence of different gauge fields within a diagram stems from the fact that the effective generators need not be traceless, so the trace that usually produces the Dynkin index no longer has to vanish for generators of different groups.  The sum of diagrams proportional to $g^2\tilde g^2$ yields a new function, $h(x,y,\tilde y)$, where $x=F/M^2$, $y=m_W^2/M^2$, $z=\wt m_W^2/M^2$, and $h(x,y,y)=f(x,y)$.  This function is given in the appendix.  The full result for the case of two $U(1)$'s is
\eqn\twous{\Delta m_0^2=\Big|\frac{F}{M}\Big|^2\sum_i\bigg[\Big(\frac{\alpha}{2\pi}\Big)^2q_i^2q_Q^2f(x_i,y)+\Big(\frac{\tilde\alpha}{2\pi}\Big)^2\tilde q_i^2\tilde q_Q^2f(x_i,\tilde y)+2\frac{\alpha\tilde\alpha}{(2\pi)^2}q_i\tilde q_iq_Q\tilde q_Qh(x,y,\tilde y)\bigg],}
where the $q$'s are the various charges of the fields in what is hopefully an obvious notation.  In general, one simply needs to transform to the mass eigenbasis and identify all of the $U(1)'s$ that result.  Each pair will have a contribution of this form.  
\centerline{\epsfxsize=2.55in\epsfbox{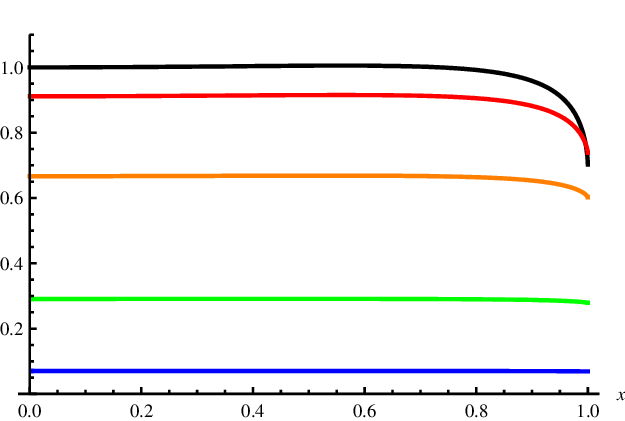}}
\centerline{Figure 4: From top to bottom, $f(x,0)$, $f(x,.1)$, $f(x,1)$, $f(x,10)$, and $f(x,100)$}
\vskip .2in
\newsec{Implications}
The mass spectrum provides some of the key predictions for the potential discovery of GMSB.  The masses calculated here (renormalized to the scale of MSSM sparticle masses \MartinZK) predict relationships among particle masses given by simple group theory factors and known functions of scales.  Our results reproduce those of standard gauge mediation \GiudiceBP\ in the appropriate limit, but provide a new set of predictions if the mediating group is Higgsed.  For example, if the messenger scale were low enough to make the mass of the $SU(2)_W$ fields non-negligible, one would find sfermions with lower than expected masses.  Additional mediating gauge fields, however, lead to higher masses.  

In ordinary gauge mediation the ratios of gaugino and sfermion masses depend primarily on the matter content of the messenger sector, but they are also highly sensitive to the gauge structure.  The modification of the spectrum can be particularly interesting if the messenger scale is near a Higgsing scale.  In that case, the massive gauge fields give significant contributions and cannot be approximated as massless (see Figure 4).  In this scenario, the ratio of gaugino and scalar masses would not readily yield the effective messenger number.  Assuming ordinary gauge mediation, one would find that it is not an integer.

Of course, a proximity of scales need not be an accident.  In \ChengYS, for example, the breaking-scale of a gauged Peccei-Quinn symmetry and the supersymmetry-breaking scale coincide.  And in the ISS model \IntriligatorDD, all scales are set by a single dimensionful parameter.
Recall that our analysis applies to the spectrum and interactions that result from \ww.  The ISS model is of this form.  It has messengers with masses, $m_\pm^2=|h\mu|^2\pm|h\mu|^2$ and a supersymmetric Higgsing of the gauge group with $m_W=g\mu$.  This gives $y=g/h$, which is naturally of order one.  The model also has $x=1$, so a small-$x$ approximation cannot be trusted. 

\bigbreak\bigskip\centerline{{\bf Acknowledgements}}\nobreak
We are grateful for the contributions of V.~Mateu, D.~Shih, and J.~Wright to this work.  K.~Intriligator deserves special thanks for suggesting the project and providing guidance along the way.  We are also indebted to N.~Craig, M.~McCullough, and J.~Thaler for pointing out that there were typos in the appendix of the first version.  Partial support for this research was provided by UCSD grant DOE-FG03-97ER40546.
\appendix{A}{The Two-Loop Calculation}
There are ten diagrams relevant to the computation of the lowest-order scalar mass correction.  They are shown in Figure 5.  The first eight are the standard contributions.  The final two arise from interactions with the scalar, $C$, of the massive vector multiplet.  For $U(1)\times U(1)'$, we have
\eqn\dterms{{\cal L}\supset-gm_WC(i\phi_+^*\phi_--i\phi_+\phi_-^*+|q|^2),}
where $\phi_\pm$ has mass-squared $M^2\pm F$, and $q$ is a scalar that will get a radiative mass.  The charge assignments should be clear.  Of course, there are many more diagrams that do not involve the messengers, but their contributions sum to zero.  In fact, Diagram 5 is independent of supersymmetry breaking, but we prefer to compute with the complete messenger multiplet.  This gives a finite result and thus a check on the calculation.  There are no IR divergences.  Dimensional reduction \SiegelWQ\ is used to regulate the UV divergences.  In this context, dimensional reduction simply amounts to performing all Lorentz algebra in four dimensions and then evaluating the resulting scalar integral in $4-2\epsilon$ dimensions.  
In evaluating the integrals, we expressed each integral as a sum of ``master integrals'' -- integrals with momentum-independent numerators. This method is discussed in more detail in \vanderBijBW.  In their notation, the most general two-loop master integral is
\eqn\gi{\vev{m_{11},m_{12},\dots|m_{21},m_{22},\dots|m_{31},m_{32},\dots}\qquad\qquad\qquad\qquad\qquad\qquad\qquad\qquad\qquad\qquad
$$
$$
\qquad\qquad\equiv\prod_{i,j,k}\int\D{k}\D{q}\frac{1}{(k^2+m_{1i}^2)(q^2+m_{2j}^2)[(k+q)^2+m_{3k}^2]}.}
In our calculation, only the following two integrals are needed, 
\eqn\threeprop{\vev{m_1|m_2|m_3}=\int\D{k}\D{q}\frac{1}{(k^2+m_1^2)(q^2+m_2^2)[(k+q)^2+m_3^2]},\quad}
\eqn\fourprop{\vev{m_1,m_1|m_2|m_3}=\int\D{k}\D{q}\frac{1}{(k^2+m_1^2)^2(q^2+m_2^2)[(k+q)^2+m_3^2]}.}  Clearly these integrals are not independent; the second is a derivative of the first.  In turns out, however, that the dimensionless integral \fourprop\ is the easier integral to evaluate, so it is useful to have the inverse identity,
\eqn\idtwo{\vev{m_1|m_2|m_3}=\frac{m_1^2\vev{m_1,m_1|m_2|m_3}+m_2^2\vev{m_2,m_2|m_3|m_1}+m_3^2\vev{m_3,m_3|m_1|m_2}}{3-d}.}
The single integral that we need is 
\eqn\singint{
\vev{m_1,m_1|m_2|m_3}\qquad\qquad\qquad\qquad\qquad\qquad\qquad\qquad\qquad\qquad\qquad\qquad\qquad\qquad\qquad\qquad
$$
$$\qquad\qquad=\frac{1}{2(4\pi)^4}\bigg[\frac{1}{\epsilon^2}+\frac{1-2\ln\bar m_1^2}{\epsilon}+1+\frac{\pi^2}{6}-2\ln\bar m_1^2+2\ln^2\bar m_1^2+2F\Big(\frac{m_2^2}{m_1^2},\frac{m_3^2}{m_1^2}\Big)\bigg],}
where $\bar m^2=m^2e^\gamma/4\pi$, and the function of the mass ratios is\footnote{$^1$}{This corrects a typo in \vanderBijBW.  Their simplified form of this function is correct.  We prefer the unsimplified form because it presents fewer numerical complications.} 
\eqnn\F
$$\eqalignno{F(a,b)
&=-\frac{1}{2}\ln^2a-\Li\Big(\frac{a-b}{a}\Big)\cr
&+\Big(\frac{a+b-1}{2r}-\frac{1}{2}\Big)\Big[\Li\Big(\frac{b-a}{x_+}\Big)-\Li\Big(\frac{a-b}{1-x_+}\Big)-\Li\Big(\frac{1-x_+}{-x_+}\Big)+\Li\Big(\frac{-x_+}{1-x_+}\Big)\Big]\cr
&-\Big(\frac{a+b-1}{2r}+\frac{1}{2}\Big)\Big[\Li\Big(\frac{b-a}{x_-}\Big)-\Li\Big(\frac{a-b}{1-x_-}\Big)-\Li\Big(\frac{1-x_-}{-x_-}\Big)+\Li\Big(\frac{-x_-}{1-x_-}\Big)\Big],\cr&&\F\cr}$$ 
having defined the parameters,
$$r=\sqrt{1-2(a+b)+(a-b)^2},\qquad x_+=\frac{1}{2}(1+b-a+r),\qquad x_-=\frac{1}{2}(1+b-a-r),$$
and having made use of the dilogarithm, 
\eqn\dilog{\Li(z)=-\int_0^1dt\frac{\ln(1-zt)}{t}.}
Finally, we have all the ingredients we need.  For the case a single supersymmetric vector superfield with mass $m_W$, the decomposition of each diagram into master integrals is shown before Figure 5.  The parameter, $\xi$, determines the gauge.  The absence of dependence on $\xi$ in the sum of diagrams provides another check on the computation.  The more general case with different vector superfields with masses $m_W$ and $\tilde m_W$ follows.  In evaluating this mixed contribution, one finds that the expressions for individual diagrams can be unwieldy when expressed in terms of two gauge-fixing parameters.  It is worth calculating with the parameters for the sake of checking the calculation, but a lot of work can be saved by adding diagrams at intermediate stages.  This has been done.  The gauge-invariant combinations are shown.

We would like to note that the vanishing of Diagram 6 is a rather robust result, though the authors know of no principle requiring it to be zero.  In particular, one can allow each gauge bosons to have arbitrary mass and to be in an arbitrary gauge.  Using four-component spinors, the diagram is found to be proportional to
\eqn\gsix{\int\D{k}\D{q}\Tr[k^\mu\Delta_{\mu\nu}(k)\gamma^\nu\Delta_{1/2}(k+q)\gamma^\rho\Delta_{1/2}(q)\Delta_{\rho\sigma}(k)k^\sigma\Delta_0(k)]}
where
\eqn\props{\Delta_{\mu\nu}(k)=\frac{-i}{k^2-m_W^2}\Big[g_{\mu\nu}-\frac{(1-\xi)k_\mu k_\nu}{k^2-\xi m_W^2}\Big],\qquad\widetilde\Delta_{\mu\nu}(k)=\frac{-i}{k^2-\widetilde m_W^2}\Big[g_{\mu\nu}-\frac{(1-\tilde\xi)k_\mu k_\nu}{k^2-\tilde\xi \widetilde m_W^2}\Big],
$$
$$
\Delta_{1/2}(k)=\frac{i(\slashed{k}+m_f)}{k^2-m_f^2},\quad\Delta_0(k)=\frac{i}{k^2}.}
A little algebra shows that \gsix\ can be written as
\eqn\gsixsimp{\int\D{k}\D{q}\frac{\Tr[\slashed{k}(\slashed{k}+\slashed{q}+m_f)\slashed{k}(\slashed{q}+m_f)]f(k^2)}{[(k+q)^2-m_f^2](q^2-m_f^2)},}
for a function, $f(k^2)$, which contains all of the information about the gauge bosons.  The rest of the integral is simplified with the use of the identity,
\eqn\num{\Tr[\slashed{k}(\slashed{k}+\slashed{q}+m_f^2)\slashed{k}(\slashed{q}+m_f^2)]=4(k\cdot q)[(k+q)^2-m_f^2]-4(k\cdot q)(q^2-m_f^2)-4k^2(q^2-m_f^2).}
If this is put back into the integral, and the change of variables, $q\rightarrow k+q$, is made in the first term, one finds that the second and third terms are exactly canceled, and the integral vanishes.

Finally, the sum of unmixed diagrams normalized so that $f(0,0)=1$ gives the function in \sm:
\eqnn\fhey
$$\eqalignno{f(x,y)
&=\frac{1}{x^2}\bigg[F(1,y)+(1+y)F\Big(\frac{1}{y},\frac{1}{y}\Big)-F(1+x,y)+\frac{1}{2}(1+x)F\Big(1,\frac{y}{1+x}\Big)\cr
&-(1+x)F\Big(\frac{1}{1+x},\frac{y}{1+x}\Big)+\frac{1}{2}(1+x)F\Big(\frac{1-x}{1+x},\frac{y}{1+x}\Big)+(x-2y)F\Big(\frac{1+x}{y},\frac{1}{y}\Big)\cr
&-\Big(1+x-\frac y2\Big)F\Big(\frac{1+x}{y},\frac{1+x}{y}\Big)+\frac{y}{2}F\Big(\frac{1+x}{y},\frac{1-x}{y}\Big)\bigg]+(x\rightarrow-x),
&\fhey\cr}$$
and the sum of the mixed diagrams gives,
\eqnn\hxyz
$$\eqalignno{h(x,y,z)
&\!=\!\bigg\{\frac{1}{2x^2(y-z)}\bigg[2(2+y)F(1,y)+(2+y)yF\Big(\frac{1}{y},\frac{1}{y}\Big)+2(x-y)F(1+x,y)\cr
&-(1+x)(4+4x-y)F\Big(1,\frac{y}{1+x}\Big)+2(1+x)(x-y)F\Big(\frac{1}{1+x},\frac{y}{1+x}\Big)\cr
&+(1+x)yF\Big(\frac{1-x}{1+x},\frac{y}{1+x}\Big)+2(x-y)yF\Big(\frac{1+x}{y},\frac{1}{y}\Big)\cr
&-(4+4x-y)\frac{y}{2}F\Big(\frac{1+x}{y},\frac{1+x}{y}\Big)\!+\frac{y^2}{2}F\Big(\frac{1+x}{y},\frac{1-x}{y}\Big)\bigg]\!\!+\!(x\rightarrow-x)\!\bigg\}\!+(y\leftrightarrow z),\cr
&
&\hxyz\cr}$$
\eject
\def\s{\vskip -.2in}
\centerline{{\bf Unmixed Diagrams}}
$$
{\rm{\bf Diagram\ 1}}\,=\,2\xi^2\langle m_+\rangle\langle m_W,m_W\rangle+2\xi^2\langle m_-\rangle\langle m_W,m_W\rangle
$$
\s
$$
{\rm{\bf Diagram\ 2}}\,=\,-2\xi^2\langle m_+\rangle\langle m_W,m_W\rangle-2\xi^2\langle m_-\rangle\langle m_W,m_W\rangle
$$
\s
$$
{\rm{\bf Diagram\ 3}}\,=\,-2(3+\xi^2)\langle m_+\rangle\langle m_W,m_W\rangle-2(3+\xi^2)\langle m_-\rangle\langle m_W,m_W\rangle
$$
\s
$$
{\rm{\bf Diagram\ 4}}\,=\,2(1+\xi^2)\langle m_+\rangle\langle m_W,m_W\rangle+2(1+\xi^2)\langle m_-\rangle\langle m_W,m_W\rangle
$$
$$
-\langle m_+|m_+|m_W\rangle-\langle m_-|m_-|m_W\rangle
$$
$$
-(4m_+^2- m_W^2)\langle m_+|m_+|m_W,m_W\rangle-(4m_-^2- m_W^2)\langle m_-|m_-|m_W,m_W\rangle
$$
\s
$$
{\rm{\bf Diagram\ 5}}\,=\,8\langle m_f\rangle\langle m_W,m_W\rangle-4\langle m_f|m_f|m_W\rangle+(8m_f^2+4m_W^2)\langle m_f|m_f|m_W,m_W\rangle
$$
\s
$$
{\rm{\bf Diagram\ 6}}\,=\,0
$$
\s
$$
{\rm{\bf Diagram\ 7}}\,=\,-2\langle m_+|m_-|0\rangle
$$
\s
$$
{\rm{\bf Diagram\ 8}}\,=\,-8\langle m_f\rangle\langle m_W ,m_W \rangle+4\langle m_+\rangle\langle m_W ,m_W \rangle+4\langle m_-\rangle\langle m_W ,m_W \rangle
$$
$$
+4\langle m_+|m_f|m_W \rangle+4\langle m_-|m_f|m_W \rangle
$$
$$
+(4m_+^2-4m_f^2-4m_W ^2 )\langle m_+|m_f|m_W ,m_W \rangle+(4m_-^2-4m_f^2-4m_W ^2 )\langle m_-|m_f|m_W ,m_W \rangle
$$
\s
$$
{\rm{\bf Diagram\ 9}}\,=\,4\langle m_+|m_-|0\rangle-4\langle m_+|m_-|m_W\rangle
$$
\s
$$
{\rm{\bf Diagram\ 10}}\,=\,-2\langle m_+|m_-|0\rangle
+2\langle m_+|m_-|m_W\rangle+2m_W^2\langle m_+|m_-|m_W,m_W\rangle
$$
\eject

\centerline{{\bf Mixed Diagrams}}
$$
{\rm{\bf Diagram\ 1}}\,+{\rm{\bf Diagram\ 3}}\,=\,-6\langle m_+\rangle\langle m_W,\wt m_W\rangle-6\langle m_-\rangle\langle m_W,\wt m_W\rangle
$$
\s
$$
{\rm{\bf Diagram\ 2}}\,+{\rm{\bf Diagram\ 4}}\,=\,2\langle m_+\rangle\langle m_W,\wt m_W\rangle+2\langle m_-\rangle\langle m_W,\wt m_W\rangle
$$
$$
-\half\langle m_+|m_+|m_W\rangle-\half\langle m_+|m_+|\wt m_W\rangle-\half\langle m_-|m_-|m_W\rangle-\half\langle m_-|m_-|\wt m_W\rangle
$$
$$
-\Big(4m_+^2-\half m_W^2-\half\wt m_W^2\Big)\langle m_+|m_+|m_W,\wt m_W\rangle-\Big(4m_-^2-\half m_W^2-\half\wt m_W^2\Big)\langle m_-|m_-|m_W,\wt m_W\rangle
$$
\s
$$
{\rm{\bf Diagram\ 5}}\,=\,8\langle m_f\rangle\langle m_W,\wt m_W\rangle-2\langle m_f|m_f|m_W\rangle-2\langle m_f|m_f|\wt m_W\rangle
$$
$$
+\Big(8m_f^2+2m_W^2+2\wt m_W^2\Big)\langle m_f|m_f|m_W,\wt m_W\rangle
$$
\s
$$
{\rm{\bf Diagram\ 6}}\,=\,0
$$
\s
$$
{\rm{\bf Diagram\ 7}}\,=\,-2\langle m_+|m_-|0\rangle
$$
\s
$$
{\rm{\bf Diagram\ 8}}\,=\,-8\langle m_f\rangle\langle m_W ,\wt m_W \rangle+4\langle m_+\rangle\langle m_W ,\wt m_W \rangle+4\langle m_-\rangle\langle m_W ,\wt m_W \rangle
$$
$$
+2\langle m_+|m_f|m_W \rangle+2\langle m_+|m_f|\wt m_W \rangle+2\langle m_-|m_f|m_W \rangle+2\langle m_-|m_f|\wt m_W \rangle
$$
$$
+(4m_+^2-4m_f^2-2m_W ^2-2\wt m_W^2 )\langle m_+|m_f|m_W ,\wt m_W \rangle
$$
$$
+(4m_-^2-4m_f^2-2m_W ^2-2\wt m_W^2)\langle m_-|m_f|m_W ,\wt m_W \rangle
$$
\s
$$
{\rm{\bf Diagram\ 9}}\,=\,4\langle m_+|m_-|0\rangle-2\langle m_+|m_-|m_W\rangle-2\langle m_+|m_-|\wt m_W\rangle
$$
\s
$$
{\rm{\bf Diagram\ 10}}\,=\,-2\langle m_+|m_-|0\rangle+\langle m_+|m_-|m_W\rangle+\langle m_+|m_-|\wt m_W\rangle
$$
$$
+m_W^2\langle m_+|m_-|m_W,\wt m_W\rangle+\wt m_W^2\langle m_+|m_-|m_W,\wt m_W\rangle
$$

\def\graphsize{1.6in}
\centertable{
\vrule height2.75ex depth0.25ex width 0pt \tabskip=1em  \hfil#\hfil&\qquad#&\hfil#\hfil\cr
\epsfxsize=\graphsize\epsfbox{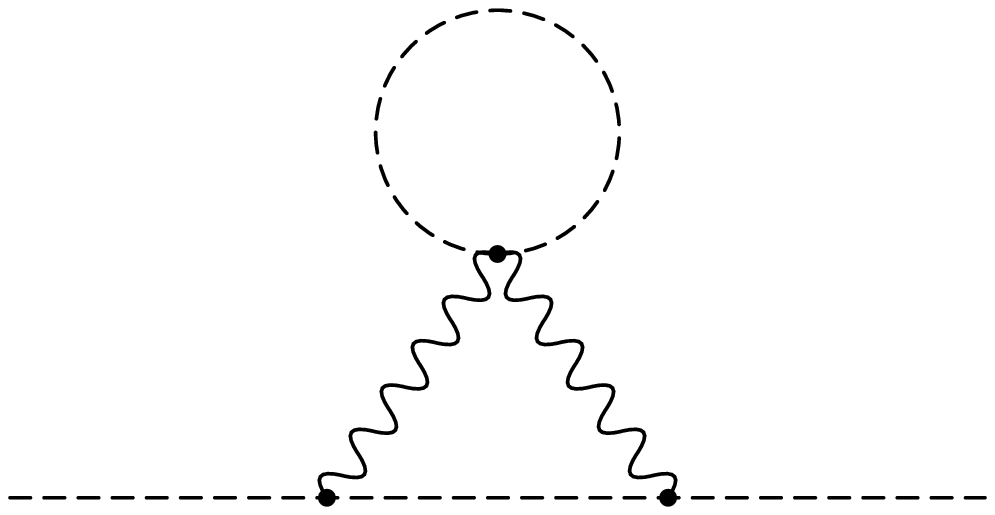}&&\epsfxsize=\graphsize\epsfbox{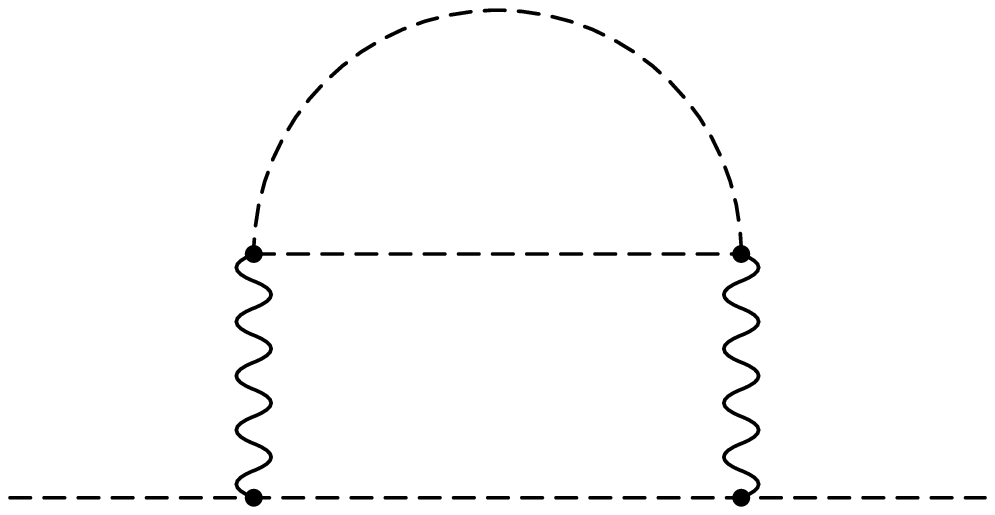}\cr
Diagram 1 && Diagram 2\cr
&&\cr
&&\cr
\epsfxsize=\graphsize\epsfbox{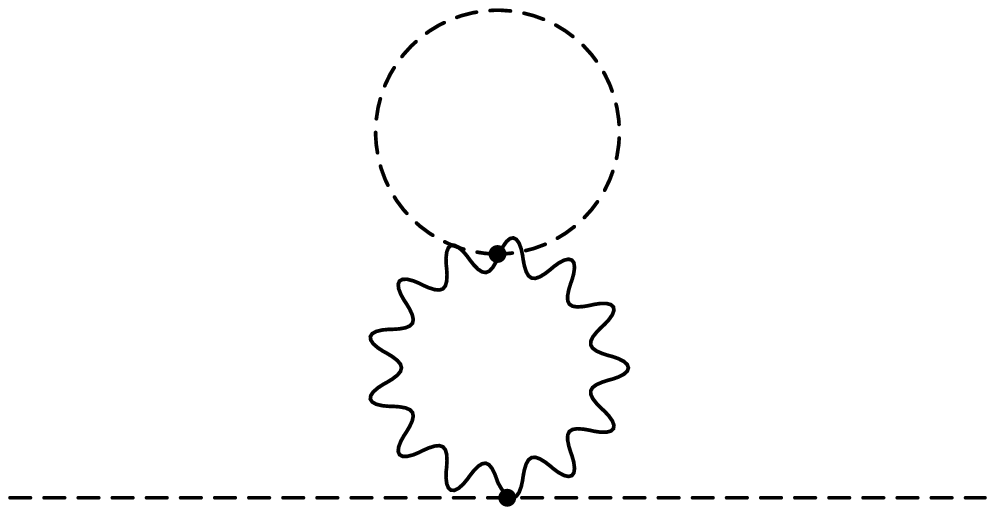}&&\epsfxsize=\graphsize\epsfbox{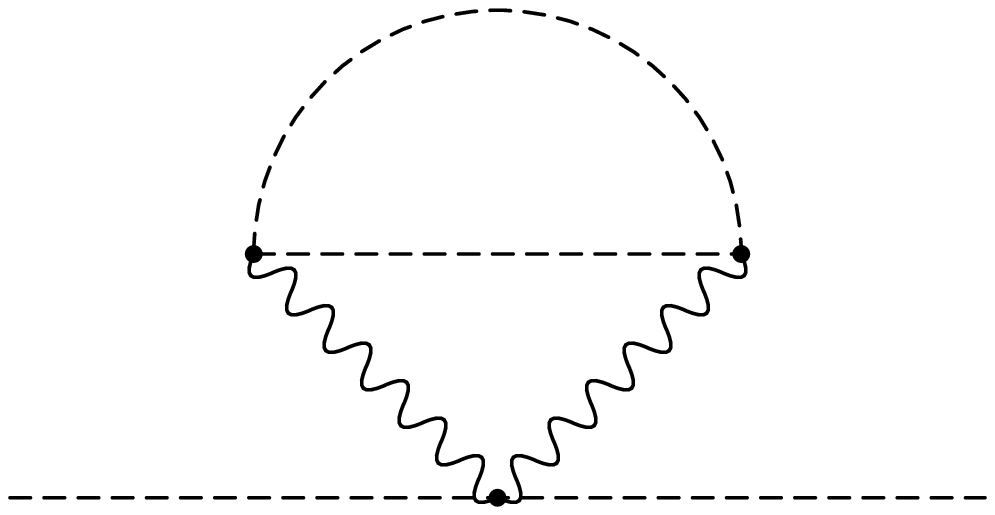}\cr
Diagram 3 && Diagram 4 \cr
&&\cr
&&\cr
\epsfxsize=\graphsize\epsfbox{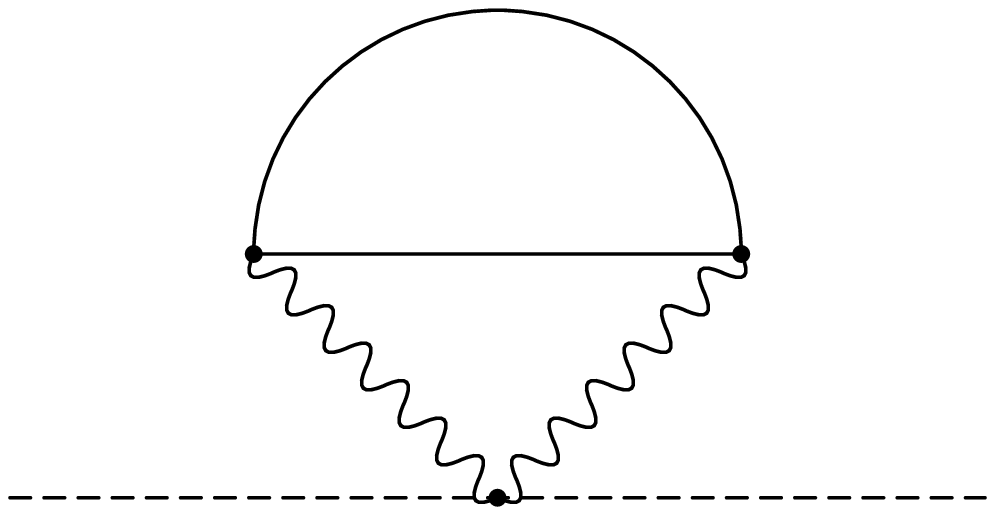}&&\epsfxsize=\graphsize\epsfbox{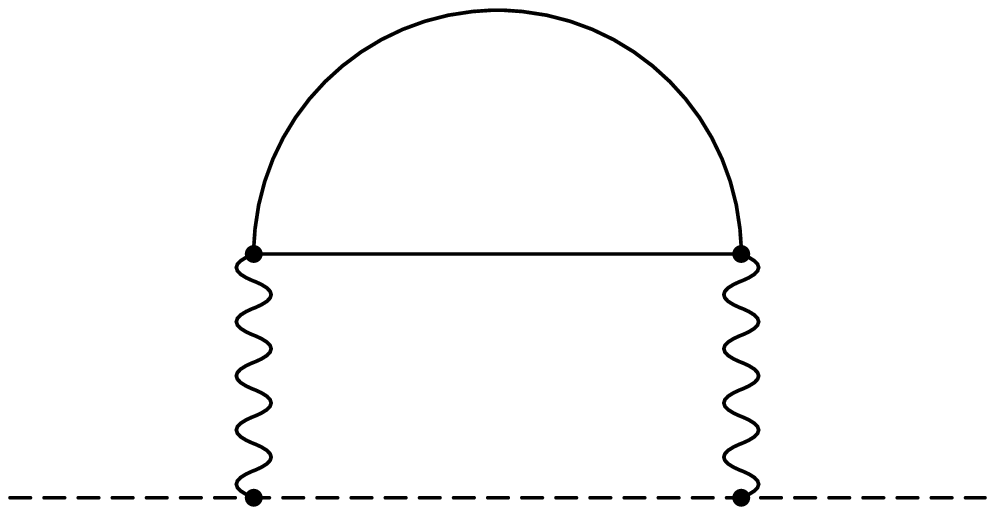}\cr
Diagram 5 && Diagram 6 \cr
&&\cr
&&\cr
\epsfxsize=\graphsize\epsfbox{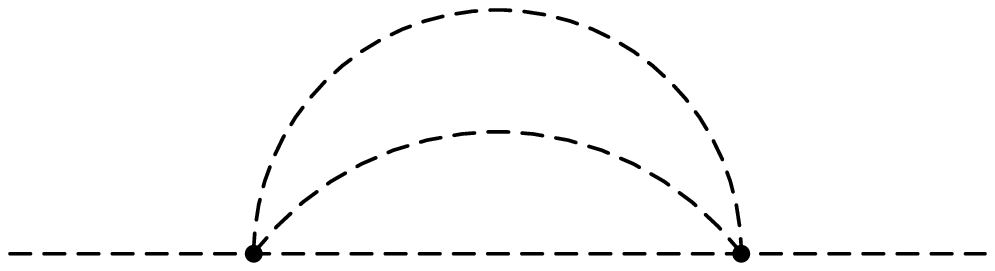}&&\epsfxsize=\graphsize\epsfbox{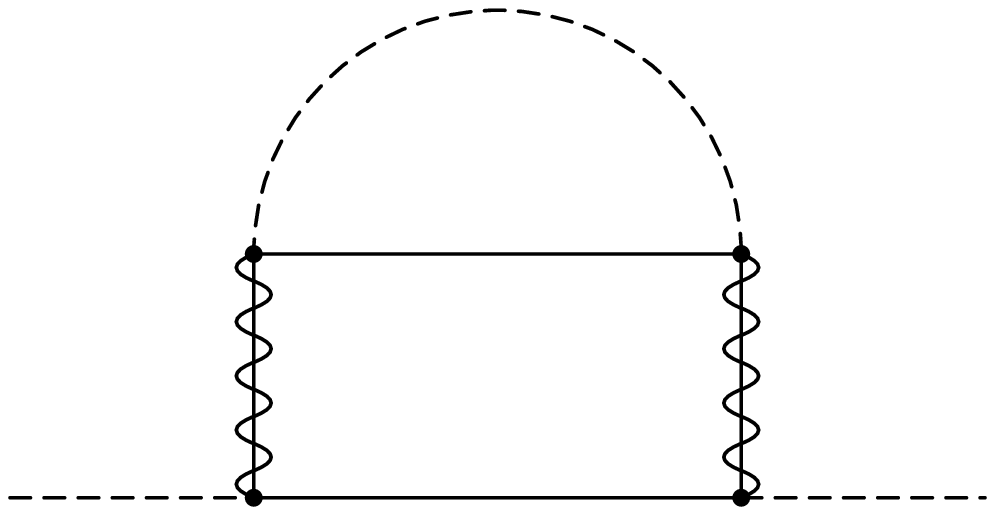}\cr
Diagram 7 && Diagram 8 \cr
&&\cr
&&\cr
\epsfxsize=\graphsize\epsfbox{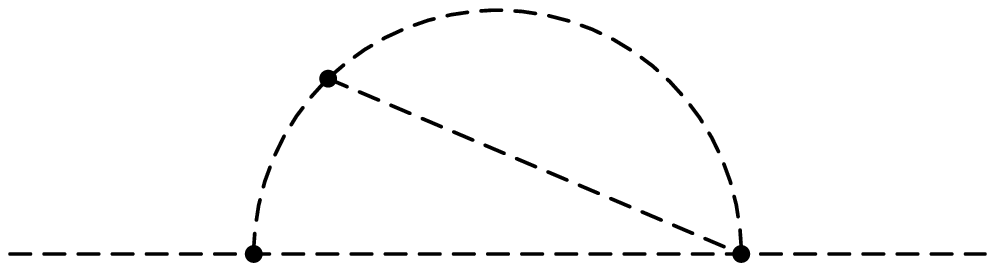}&&\epsfxsize=\graphsize\epsfbox{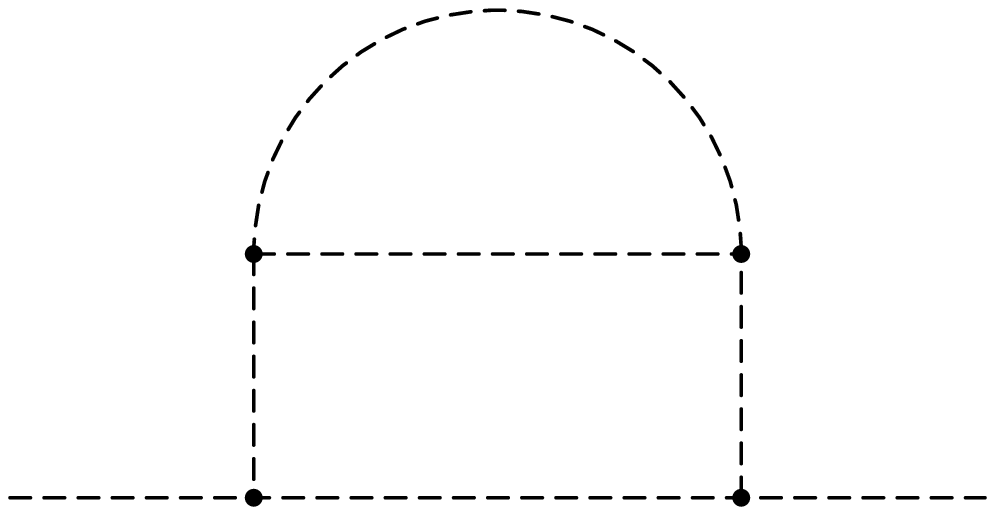}\cr
Diagram 9 && Diagram 10 \cr
}
\centerline{Figure 5: The two-loop diagrams contributing to MSSM scalar masses}
\vskip .3in
\eject
\listrefs
\bye